\documentclass[useAMS,usenatbib]{mn2e}
\usepackage{graphicx,amsmath}

\title{RXTE-PCA Observations of XMMU J054134.7$-$682550}
\author[S.\c{C}. Inam et al.]{S.\c{C}. \.{I}nam$^1$\thanks{inam$@$baskent.edu.tr}, L.J. Townsend$^2$, V.A. McBride$^2$, A. Baykal$^3$\thanks{altan$@$astroa.physics.metu.edu.tr}, M.J. Coe$^2$, R.H.D. Corbet$^4$ \\
$^1$ Department of Electrical and
Electronics Engineering, Ba\c{s}kent University, 06530 Ankara, Turkey  \\
$^2$ School of Physics and Astronomy, Southampton University, SO17 1BJ, UK\\
$^3$ Physics Department, Middle East Technical University, 06531 Ankara,
Turkey \\  
$^4$ University of Maryland, Baltimore County, Mail Code 662, NASA Goddard Space Flight Center, Greenbelt, MD 20771, USA}

\date{}
\pagerange{\pageref{firstpage}--\pageref{lastpage}} \pubyear{2002}

\begin{document}
\label{firstpage}

\maketitle

\begin{abstract}
We analyzed RXTE-PCA observations of a recent outburst of the X-ray pulsar XMMU~J054134.7$-$682550. We  calculated the pulse frequency history of the source. We found no sign of a binary companion.  The source spins up when the X-ray flux is higher, with a correlation between the spin-up rate and X-ray flux, which may be interpreted as a sign of an accretion disk. On the other hand, the source was found to have an almost constant spin frequency when the X-ray flux is lower without any clear sign of a spin-down episode. The decrease in pulsed fraction with decreasing X-ray flux was intrepreted as a sign of accretion geometry change, but we did not find any evidence of a transition from accretor to propeller regimes. The source was found to have variable pulse profiles. Two peaks in pulse profiles were usually observed. We studied the X-ray spectral evolution of the source throughout the observation. Pulse phase resolved analysis does not provide any further evidence for a cyclotron line, but may suggest a slight variation of intensity and width of the 6.4\,keV iron line with phase.       
\end{abstract}


\section{Introduction}

The X-ray pulsar XMMU J054134.7$-$682550 in the Large Magellanic Cloud (LMC) was
discovered as a High Mass X-ray Binary (HMXB) in an XMM-Newton survey
of the LMC (Shtykovskiy \& Gilfanov 2005). The source was then listed as an HMXB in
a catalog of HMXBs in the Magellanic Clouds (Liu, van Paradijs \& van den Heuvel 
2005).

From a routine scanning of Swift-BAT data on 3 August 2007, XMMU J054134.7$-$682550 was found to be in outburst with an increase in flux levels from $\sim 10-20$ mCrab to $\sim 50$ mCrab (Palmer, Grupe \& Krimm 2007).  Spectral features, the transient nature, and pulsations at 61.6\,s detected in RXTE observations during the outburst were considered to be an indication of the high mass X-ray binary nature of the source (Markwardt, Swank \& Corbet 2007).  These authors also noted cyclotron absorption features at $\sim 10$ and $\sim 20$\,keV in the X-ray spectrum of the source. 

In this paper we investigate the X-ray timing and spectral properties of the recent outburst of XMMU~J054134.7$-$682550 from 9 August 2007 through 1 November 2007 (MJD 54321 -- MJD 54405). During this period  the 3--20\,keV flux declined from $\sim 3.3\times 10^{-10}$\,erg\,s$^{-1}$\,cm$^{-2}$ to $\sim 6\times 10^{-12}$ \,erg\,s$^{-1}$\,cm$^{-2}$.  We track the changes in spin period of the source, relating these to changes in the accretion geometry, and search for signatures of binarity in the lightcurve.  We also study the evolution and flux dependence of the pulsed fraction through the outburst.  Furthermore, we explore the spectral evolution of the source over the outburst and with neutron star spin phase and investigate the possibility of cyclotron features in the spectrum.

\section{Instruments and Observations}

The Proportional Counter Array (PCA, Jahoda et al. 2006) on board the Rossi X-ray Timing Explorer (RXTE) consists of five identical
proportional counters coaligned to the same point in the sky, with a total geometric area of approximately 6250\,cm$^2$, a field of view to 1 square degree, and operating in the 2--60\,keV energy range. In May 2000 and in December 2006, PCU0 and PCU1, respectively, lost their propane veto layers. Without a propane veto layer, these PCUs have higher background levels. We have excluded the data from these PCUs in our spectral analysis. To obtain better energy channel counting statistics, we used only the data obtained from the top PCU layers for the spectral analysis. For both spectral and timing analysis, the latest combined background models were used together with HEASOFT 6.3 to estimate the appropriate background.  

\begin{figure}
\includegraphics[width=80mm,angle=-90]{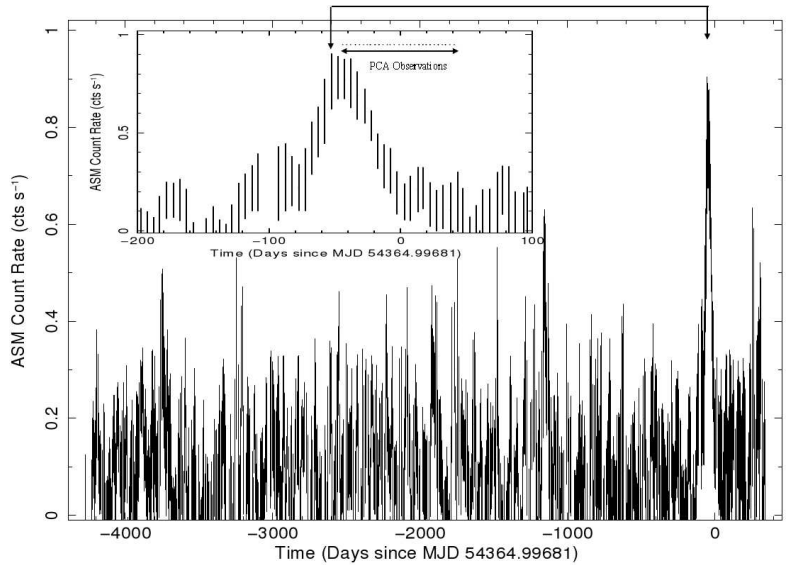}
\caption{Long-term RXTE-ASM Lightcurve of XMMU~J054134.7$-$682550. The inset shows a $\sim300$\,day portion around the outburst at $\sim$ MJD 54365. Dots in the inset represent individual PCA observations listed in Table~\ref{tab:obs}.}
\label{fig:asm}
\end{figure}

\begin{figure}
\includegraphics[width=75mm]{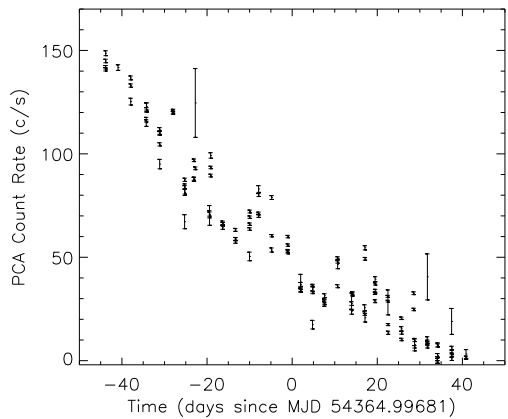}
\caption{61\,s binned 3--25 keV lightcurve of XMMU~J054134.7$-$682550}
\label{fig:lightcurve}
\end{figure}

The long-term RXTE-ASM lightcurve of the source is presented in Fig.~\ref{fig:asm}. RXTE observed XMMU~J054134.7$-$682550 through a Type II outburst and 28 observations were undertaken between 9 August 2007 and 1 November 2007 with a total exposure of $\sim 127$\,ks (see Table~\ref{tab:obs}). A 3--25\,keV background subtracted RXTE PCA lightcurve of the source is presented in Fig.~\ref{fig:lightcurve}.  Although the number of active PCUs varied between one and three during the observations, count rates in Fig.~\ref{fig:lightcurve} have been adjusted as if 5 PCUs were active using the {\tt correctlc} tool in HEASOFT.

\begin{table}
\caption{PCA observations of XMMU~J054134.7$-$682550.  The MJD column refers to the start of the observation.}
\begin{tabular}{l l r}
\hline
Obs ID & Time & Exposure  \\
93413-01- & (MJD) & (ks)  \\
\hline
01-00 & 54321.108 & 3.5 \\
03-00 & 54327.014 & 2.0  \\
04-00 & 54330.669 & 1.4  \\
04-01 & 54330.733  & 1.7 \\
05-00 & 54333.796 & 3.9  \\
06-00 & 54336.954 & 2.5  \\
07-00 & 54339.707 & 3.2  \\
08-01 & 54341.875 & 2.1  \\
08-00 & 54342.197 & 0.9  \\
09-00 & 54345.526 & 1.4  \\
09-01 & 54345.798 & 2.0  \\
10-00 & 54348.613 & 1.4  \\
10-01 & 54348.678 & 1.6  \\
11-00 & 54351.595 & 3.8  \\
12-00 & 54354.950 & 5.5  \\
13-00 & 54357.048 & 3.1  \\
14-00 & 54360.133 & 1.8  \\
15-00 & 54363.926 & 3.1  \\
16-00 & 54366.932 & 2.9  \\
17-00 & 54369.753 & 2.3 \\
18-00 & 54372.491 & 1.4 \\
18-01 & 54372.560 & 1.5  \\
19-00 & 54375.628 & 2.6  \\
20-00 & 54378.914 & 1.6  \\
20-01 & 54379.121 & 1.4  \\
21-00 & 54381.995 & 1.4  \\
21-01 & 54382.06 & 1.5  \\
22-00 & 54384.42 & 1.3  \\
22-01 & 54384.474 & 1.6  \\
23-00 & 54387.420 & 1.5  \\
23-01 & 54387.486 & 1.5  \\
24-00 & 54390.619 & 3.2  \\
25-01 & 54393.519 & 1.5  \\
25-00 & 54393.712 & 2.1  \\
26-00 & 54396.704 & 3.5  \\
27-00 & 54399.067 & 1.6  \\
27-01 & 54399.143 & 2.5  \\
28-00 & 54402.451 & 3.8  \\
29-00 & 54405.807 & 2.2  \\
\hline
\end{tabular}
\label{tab:obs}
\end{table}

\section{Timing Analysis}

\begin{figure*}
\includegraphics[width=120mm]{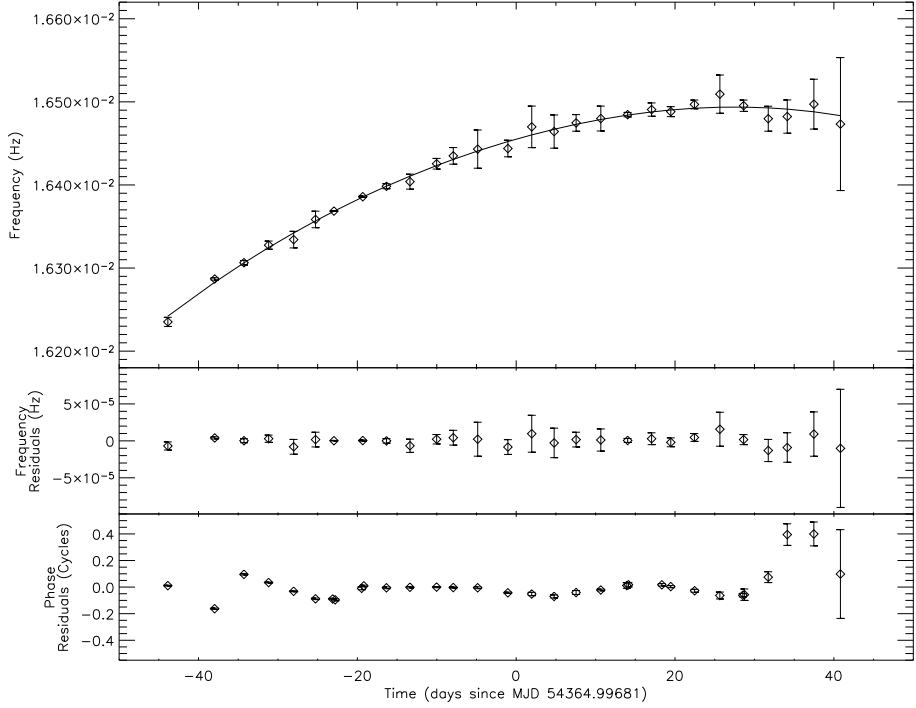}
\label{fig:frequency}
\caption{Pulse frequency and its best fit model (solid line), pulse frequency residuals and pulse phase residuals time series of XMMU~J054134.7$-$682550. Each measurement corresponds to a single observation.}
\end{figure*}

We used background corrected lightcurves for timing analysis.
 These lightcurves were also corrected to
 the barycentre of the Solar system. From the periodogram around 61\,s, a period was found giving the maximum $\chi^2$. A template pulse profile from each RXTE observation was constructed by folding the data on this period. The template pulse profile comprising 20 phase bins was analytically represented
 by its Fourier harmonics (Deeter \&
Boynton 1985) and cross-correlated with the harmonic representation
of average pulse profiles obtained from each observation.
The maximum value of the cross-correlation is well-defined and does not depend on the phase binning of the pulses. We used
a 10-term unweighted harmonic series to cross-correlate the template
 pulse profile with the pulse profiles obtained from $\sim 400-500$\,s
 long segments of the observation.
 For each single or two succeeding RXTE observations, we obtained $\sim 5-12$ arrival times. From the slopes of arrival times
($\delta\phi=\phi_0+\delta\nu(t-t_0)$, where $\delta\phi$ is the pulse phase offset deduced from the pulse timing
analysis, $t_0$ is the mid-time of the observation, $\phi_0$ is the phase offset
at $t_0$ and $\delta\nu$ is the deviation from the mean pulse frequency),
we corrected the pulse frequencies which are presented in Fig.~\ref{fig:frequency}
(top panel).

We were able to phase connect all observation segments by using a fifth-degree polynomial in a Taylor expansion,
\begin{align}
\delta\phi = & \phi_0+\delta\nu(t-t_0)+{{1} \over {2}}\dot{\nu}(t-t_0)^2+
{{1} \over {6}}\ddot{\nu}(t-t_0)^3 \nonumber \\
                 & +{{1} \over {24}}\dddot{\nu}(t-t_0)^4
+{{1} \over {120}}\ddddot{\nu}(t-t_0)^5
\end{align}
where   $\dot{\nu}$, $\ddot{\nu}$, $\dddot{\nu}$, $\ddddot{\nu}$ are the
 first, second, third and fourth derivatives of the pulse frequency of the source. Pulse residuals from the subraction of the fifth degree polynomial are presented in Fig.~\ref{fig:frequency}.

\begin{table}
\caption{Timing Solution of XMMU J054134.7$-$682550}
\begin{tabular}{l | c} \hline
Parameter &  Value  \\ \hline
Epoch & MJD 54364.99681(4) \\
Spin Frequency($\nu$) & $1.6455121(7)\times 10^{-2}$ Hz \\
Spin Frequency Derivative ($\dot \nu$) & $3.1412(1)\times 10^{-11}$ Hz s$^{-1}$ \\
$\ddot \nu$ & $-1.2274(7)\times 10^{-17}$ Hz s$^{-2}$ \\
$\dddot \nu$ & $-1.25(8)\times 10^{-25}$ Hz s$^{-3}$ \\
$\ddddot \nu$ & $-8.0(1)\times 10^{-31}$ Hz s$^{-4}$ \\
RMS Residual (pulse phase) & 0.056
\\ \hline
\end{tabular}
\label{tab:timing}
\end{table}

\begin{figure*}
%\begin{tabular}{c c}
\begin{minipage}{65mm}
\includegraphics[width=6.5cm]{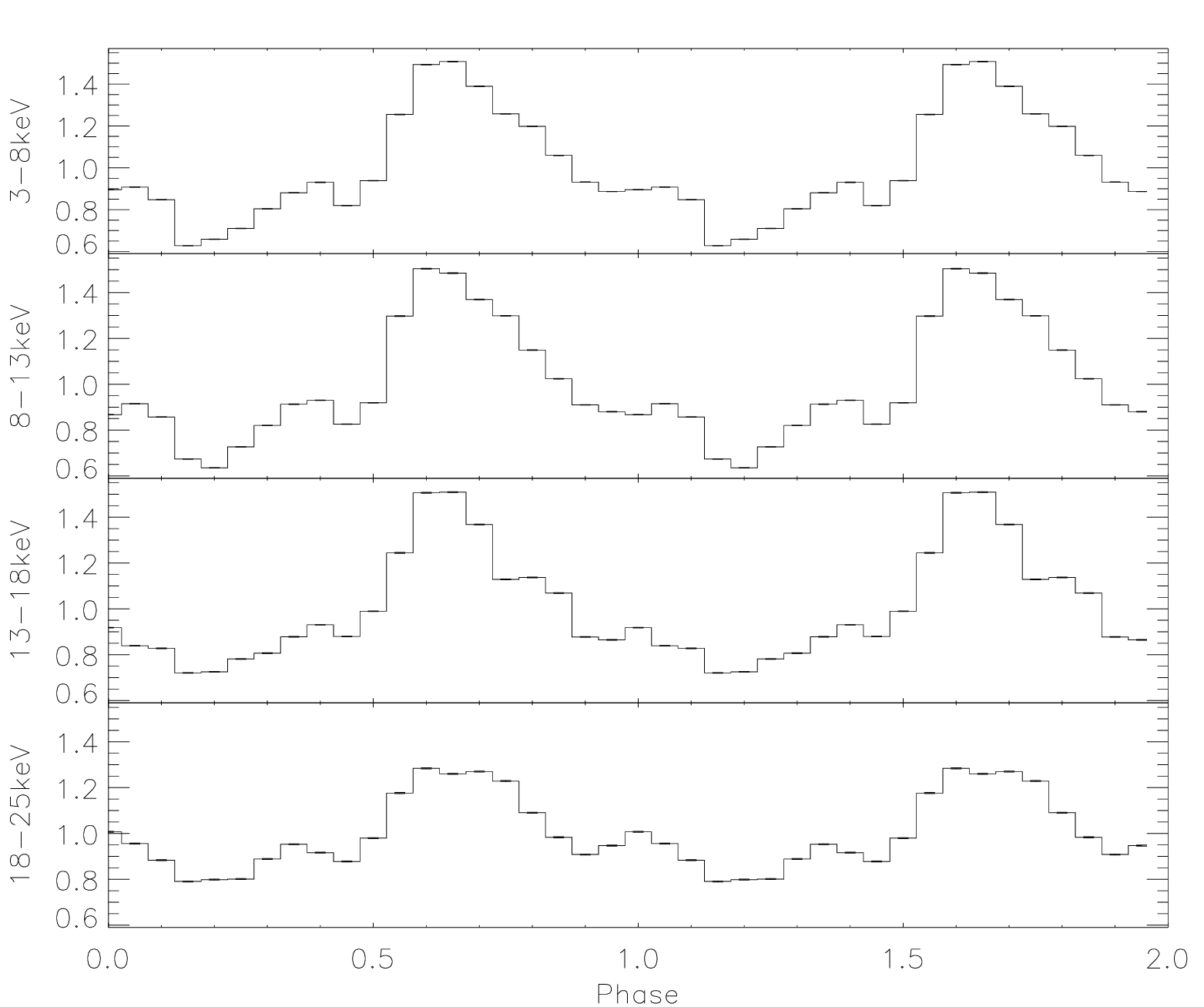}
\end{minipage} 
\hspace{0.5cm}
\begin{minipage}{65mm}
\includegraphics[width=6.5cm]{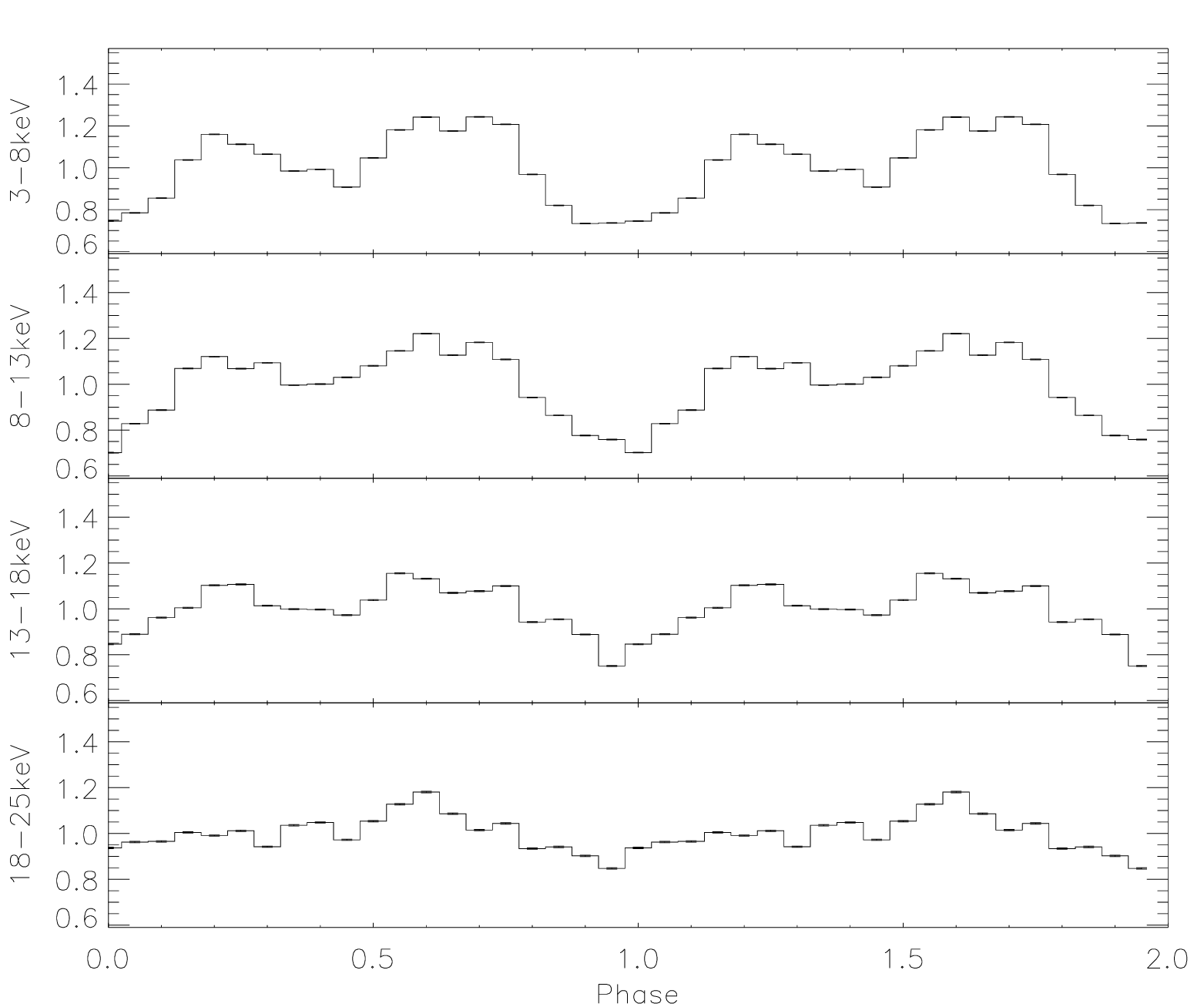} 
\end{minipage}\\
\vspace{0.5cm}
\begin{center}
\begin{minipage}{65mm}
%\hspace{5cm}
\includegraphics[width=6.5cm]{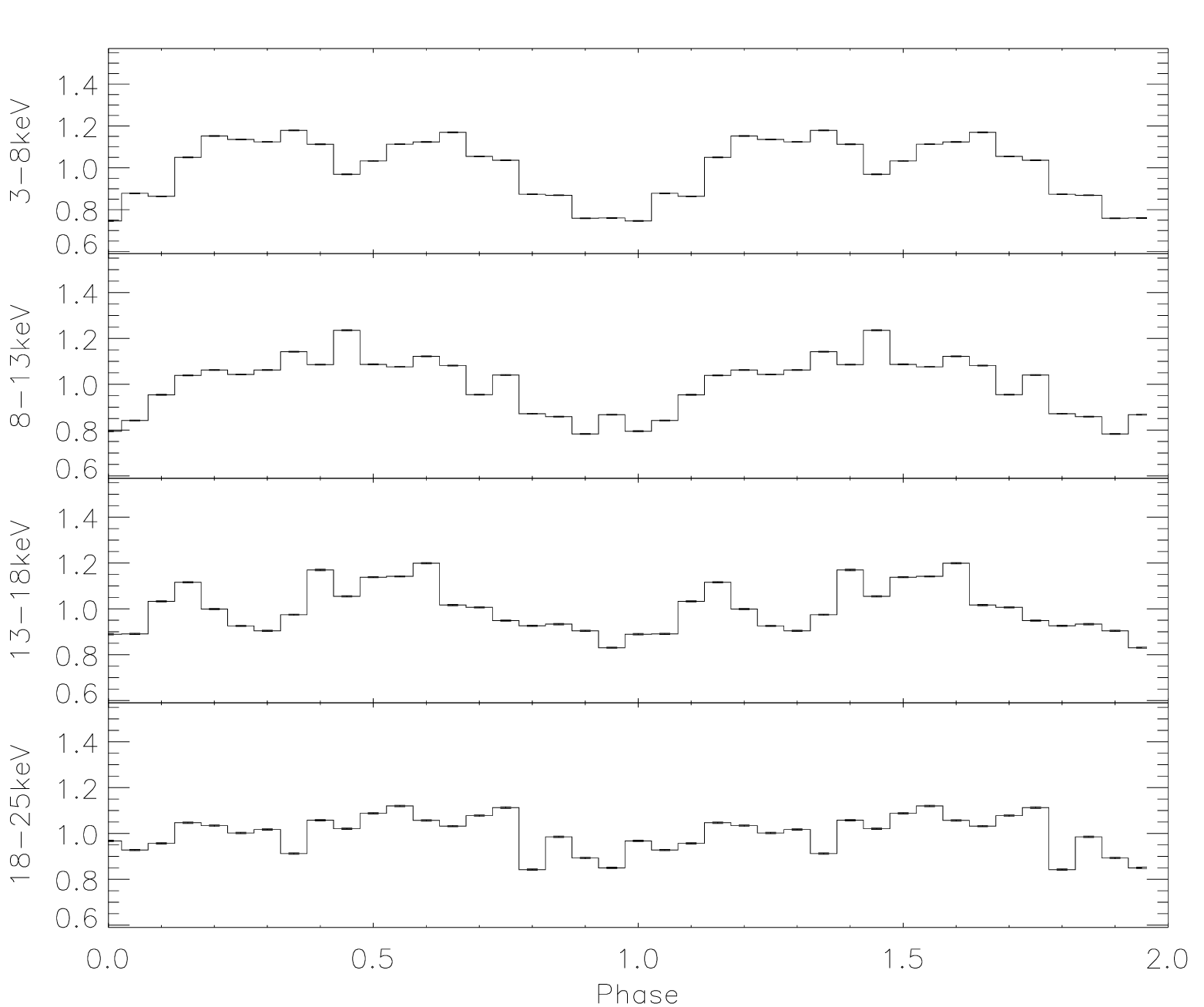} 
\end{minipage}
\end{center}
\caption{Energy dependent pulse profiles for observations on MJD 54321.1 (top left panel), MJD 54351.6 (top right panel), and MJD 54387.4 (bottom panel). The unit of y-axis of each plot is normalized count from the specified energy interval.}
\label{fig:profiles}
\end{figure*}

Table~\ref{tab:timing} presents the timing solution of XMMU~J054134.7$-$682550.
 In the top panel of Fig.~\ref{fig:frequency}, we present independent
pulse frequency measurements for each observation and
 best fit model given by
timing solution. In the middle and bottom panels, pulse frequency
 and pulse phase residuals are presented. In both residual panels, we have not seen any signature of a Doppler shift due to
 binary orbital modulation.

\begin{figure*}
%\begin{tabular}{c c}
\begin{minipage}{75mm}
\includegraphics[width=65mm]{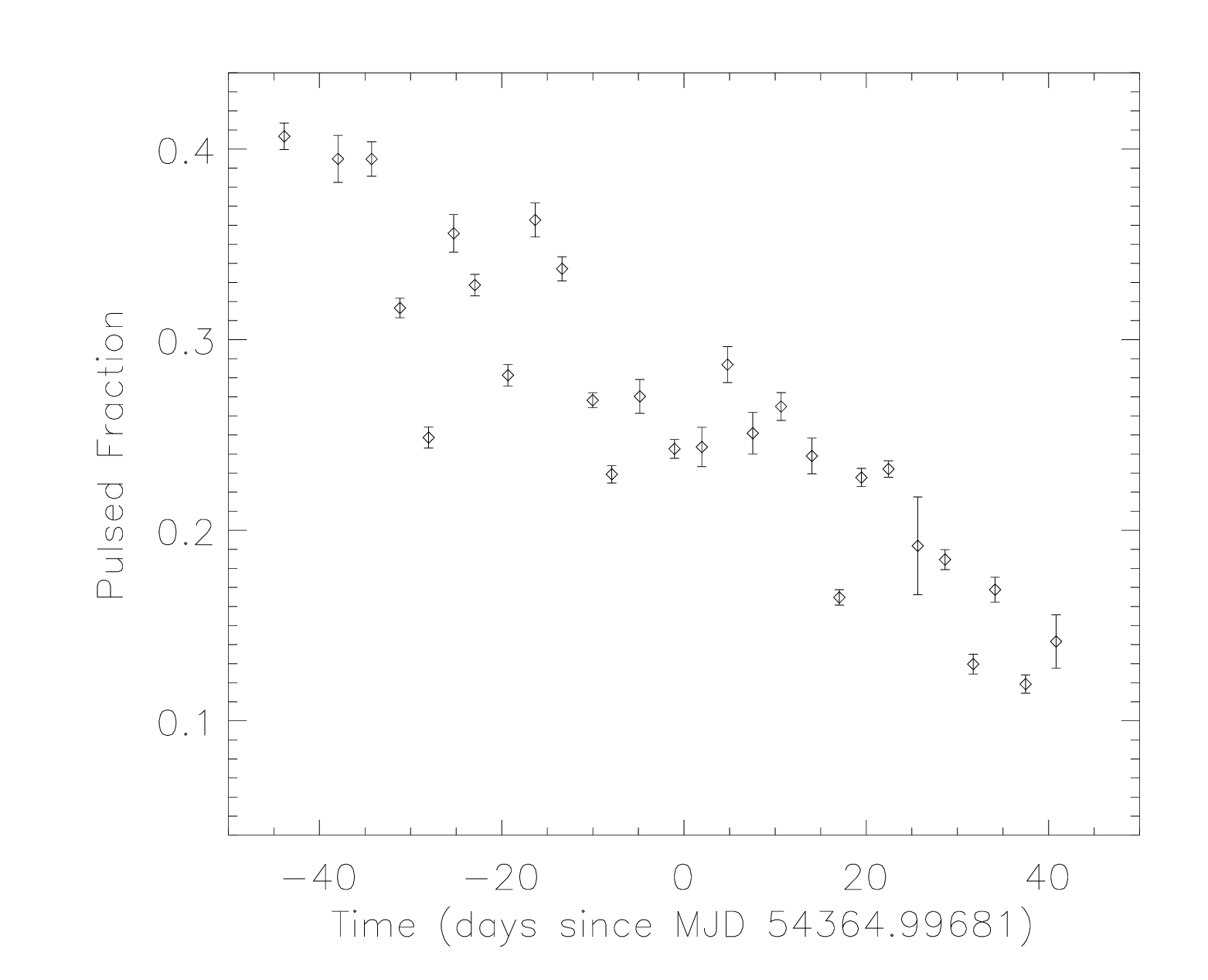}
\end{minipage}
\hspace{5mm}
\begin{minipage}{75mm}
\includegraphics[width=65mm]{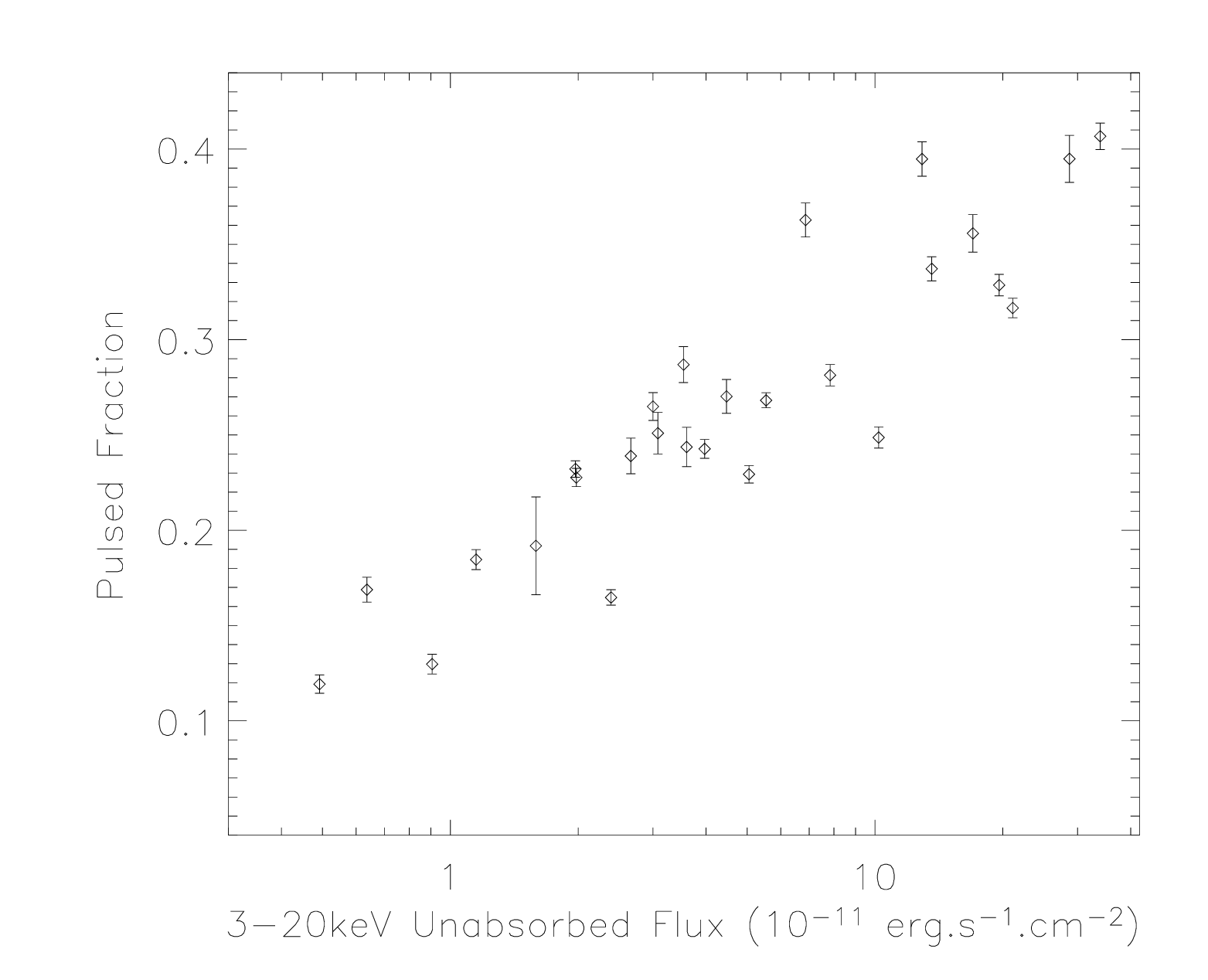}
\end{minipage}
%\end{tabular} 
%\end{center}
\caption{Evolution and X-ray flux dependence of pulsed fraction for XMMU~J054134.7$-$682550.}
\label{fig:pfrac}
\end{figure*}

Folding 3--8\,keV, 8--13\,keV, 13--18\,keV and 18--25\,keV light-curves of each observation with the pulse frequencies presented in Fig.~\ref{fig:frequency}, we obtained energy dependent pulse profiles of the source. We present 12 sample pulse profiles from three observations in Fig.~\ref{fig:profiles}. We also obtained pulse profiles from 3--25\,keV light-curves. Using these pulse profiles, we calculated pulsed fractions using the definition $(F_{\rm max}-F_{\rm min})/(F_{\rm max}+F_{\rm min}))$ where
$F_{\rm max}$ and $F_{\rm min}$ are the highest and lowest fluxes of the phase bins. In Fig.~\ref{fig:pfrac}, we present evolution and X-ray flux dependence of the pulsed fraction of the source. X-ray flux values are obtained using the spectral analysis of the source (see next section). 

\begin{figure}
%\begin{center}
%\hspace{1.1cm}
\includegraphics[width=75mm]{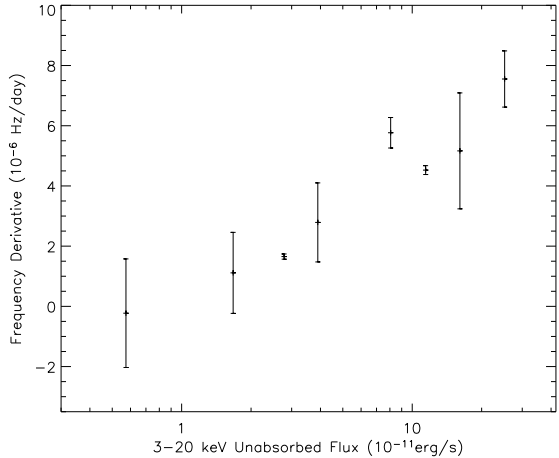} 
%\end{center}
%\begin{center}
\caption{Variation of frequency derivative with X-ray flux.}
\label{fig:fderiv}
%\end{center}
\end{figure}

For each sequence of 3--4 pulse frequency measurements, we fit a straight line to the pulse frequency time series in order to find the average pulse frequency derivatives. We plot the frequency derivative as a function of the average 3--20\, keV X-ray flux in Fig.~\ref{fig:fderiv}.  

\section{Spectral Analysis}

Spectrum, background and response files for the observations were
created using HEASOFT 6.3.  We did not use channels corresponding to
photon energies below 3\,keV due to uncertainties in the background at
these energies; neither did we use channels above 20\,keV (25\,keV for the brightest observation on MJD 54321) due to low
count statistics.  No systematic error was included in the fits, as statistical errors dominate for these data.

\begin{figure}
%\begin{center}
\includegraphics[width=85mm]{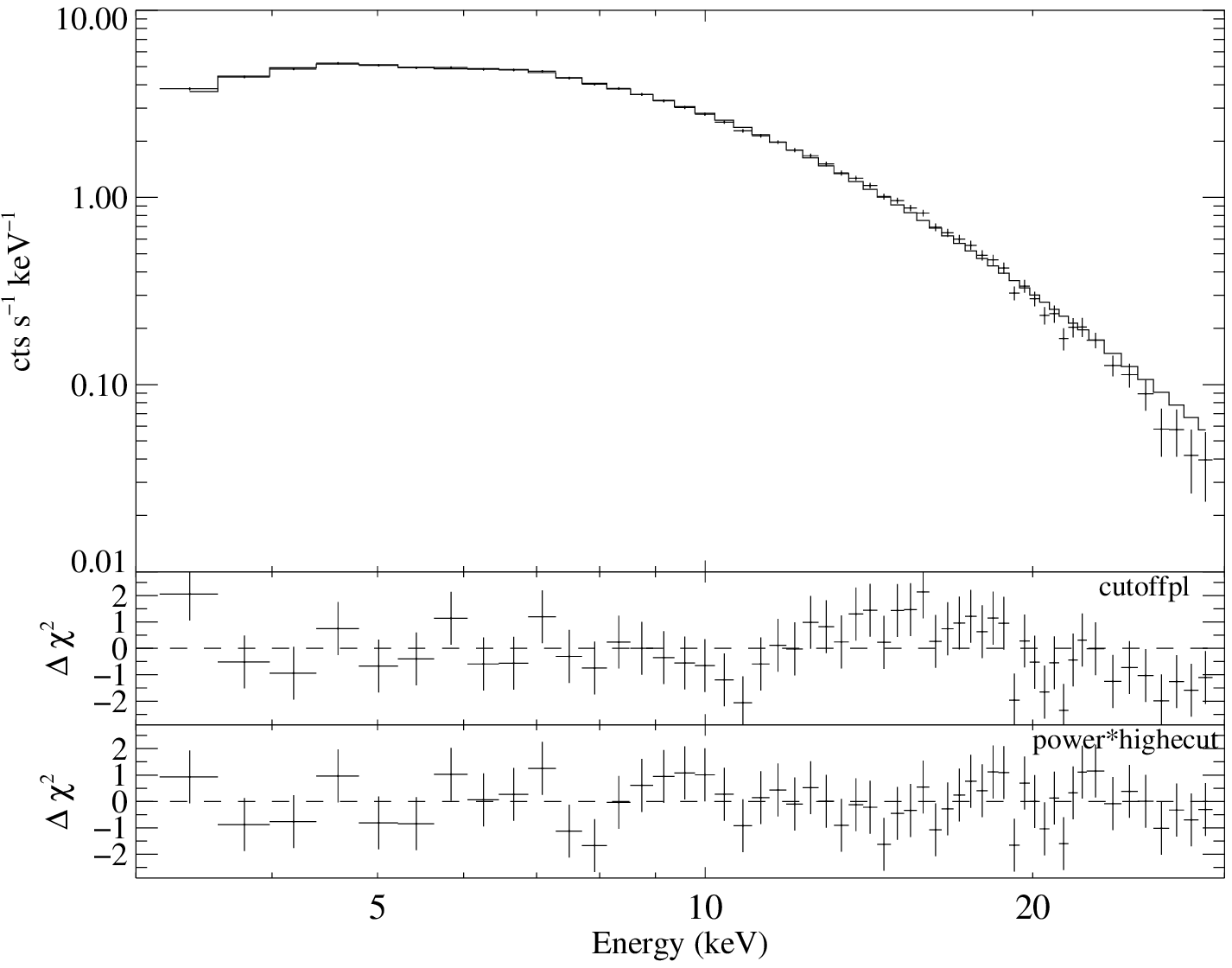}
\caption{The upper panel shows the cutoff power law model fitted to the spectrum. The middle panel shows the $\Delta\chi^2$ residuals from this model fit. A dip at around 10\,keV is clearly shown, but the model is a poor fit to the continuum in general. The lower panel shows the $\Delta\chi^2$ residuals from fitting a power law model with a high energy cutoff. The dip at 10\,keV is no longer apparent, showing that a cyclotron component at 10\,keV is not necessary.}
\label{fig:phaseavg}
%\end{center}
\end{figure}

The brightest observation, on MJD 54321, was fit with an absorbed cut-off power law  (White, Swank \& Holt 1983) and a Gaussian emission component to
account for a weak iron line at 6.4\,keV.  The neutral hydrogen density was unconstrained by the fit, and thus fixed to the line-of-sight value of $2.5\times10^{21}$\,cm$^{-2}$ (Kalberla et al. 2005) .  The fit, along with $\Delta\chi^2$ residuals is shown in Fig.~\ref{fig:phaseavg} and reported in column 2 of Table~\ref{tab:spec}.  

Given that this model does not provide an optimal description of the observed spectrum ($\chi^2_\nu=1.7$), and that
Markwardt et al. (2007) reported possible cyclotron line features at 10 and 20\,keV, we introduced cyclotron absorption features into the model at these energies.  The model fit was clearly improved, however, the F-test probability of this improvement being by
chance is quite significant:  0.018.  

An alternative continuum model was then used ({ \tt powerlaw * highecut}) which provided a better fit (see Table~\ref{tab:spec} and residuals in Fig.~\ref{fig:phaseavg}) of the continuum without the need for a cyclotron feature (see Fig.~\ref{fig:phaseavg}),  lower panel).  In this fit we have allowed the neutral hydrogen density to vary, but it is still poorly constrained by the model. This model introduces a discontinuity in the spectrum at the cutoff energy, and such a discontinuity can disguise the presence of a cyclotron feature.  Thus, our spectral analysis cannot confirm or rule out a cyclotron line in this source.

\begin{table}
\caption{Results of spectral analysis.  Errors are 90\% uncertainties. Flux is in erg\,cm$^{-2}$\,s$^{-1}$ and the Fe normalisation in photons\,cm$^{-2}$\,$s^{-1}$.}
\begin{tabular}{l r r}
\hline
 Parameter & {\tt cutoffpl} & {\tt power * highecut}\\
 \hline
 \smallskip
 $N_{\rm H}\times10^{21}$\,cm$^{-2}$ & 2.5 (fixed) & 0.8$^{+0.8}_{-0.8}$\\
 \smallskip
 $\Gamma$ & 0.42$^{+0.05}_{-0.05}$ & 0.95$^{+0.05}_{-0.09}$\\
 \smallskip
 $E_{\rm cut}$  (keV) & -- & 14.7$^{+0.8}_{-0.8}$\\
 \smallskip
 $E_{\rm fold}$ (keV) & 14.3$^{+0.5}_{-1.1}$ &15$^{+1}_{-1}$\\
 \smallskip
 $E_{\rm Fe}$ (keV)& 6.5$^{+0.2}_{-0.2}$ & 6.5$^{+0.3}_{-0.4}$\\
 \smallskip
 $\sigma_{\rm Fe}$ (keV) & 0.3$^{+0.3}_{-0.3}$ & 1.0$^{+0.5}_{-0.4}$\\
 \smallskip
 Fe normalisation & 2.4$^{+0.9}_{-0.8}\times10^{-4}$ & 7$^{+5}_{-3}\times10^{-4}$\\
 \smallskip
 Unabs Flux (3--25\,keV) & $4.2\times10^{-10}$ & $4.2\times10^{-10}$\\
 \smallskip
 $\chi^2_\nu$ (d.o.f) & 1.7 (50) & 0.72 (48) \\ 
 \hline
\end{tabular}
\label{tab:spec}
\end{table}

%\begin{center}
%\small{Table 2.  }
%\begin{tabular}{| l | c | c | c |} 
%\hline
%Parameter  &  Obs. 1 & Obs 2 & Obs. 3  \\ \hline
%Time (MJD) & 54321.135 & 54351.623 & 54387.435 \\
%$N_H$ ($10^{22}.cm^{-2}$) & $0.51\pm 0.50$ & $2.65\pm 0.51$ & $2.94\pm 2.24$ \\
%Power Law Index & $0.91\pm 0.05$ & $0.95\pm 0.01$ & $1.08\pm 0.16$ \\
%Power Law Normalization ($10^{-3}.$cts.cm$^{-2}.$s$^{-1}$) & $7.58\pm 1.12$ &
%$3.83\pm 0.25$ & $1.13\pm 0.75$ \\
%Cut-off Energy (keV) & $15.3\pm 0.5$ & $19.1\pm 0.9$ & $11.5\pm 2.8$ \\
%Folding Energy (keV) & $18.5\pm 1.9$ & $7.0\pm 2.3$ & $24.3\pm 12.9$ \\
%Iron Line Energy (keV) & $6.36\pm 0.18$ & $6.59\pm 0.25$ & - \\
%Iron Line Sigma (keV) & $0.92\pm 0.26$ & $0.41\pm 0.33$ & - \\ 
%Iron Line Normalization ($10^{-4}.$cts.cm$^{-2}.$s$^{-1}$) & $3.95\pm 2.86$ &
%$0.59\pm 0.35$ & - \\
%3-25keV Unabsorbed Flux ($10^{-11}.$erg.cm3-25$^{-2}.$s$^{-1}$) & $30.5\pm 4.5$ & 
%$14.0\pm 7.6$ & $2.8\pm 1.9$ \\
%Reduced $\chi^2$ (40 d.o.f. for Obs.1 and 2, 43 d.o.f. for Obs.3) & 0.64 & 1.02 & 0.63 
%\\ \hline
%\end{tabular}}
%\end{center}
%\end{table}  

\begin{figure}
%\begin{center}
\includegraphics[width=75mm]{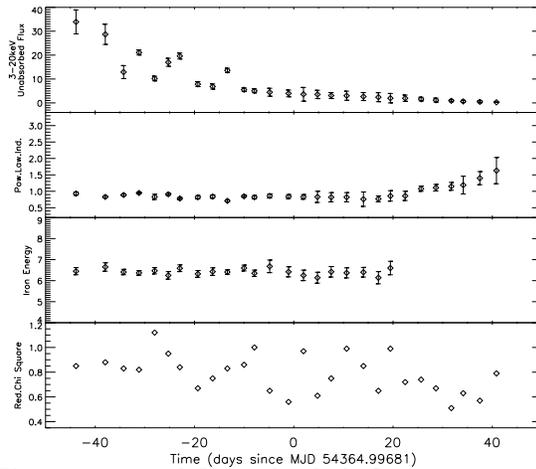} 
%\end{center}
\caption{3--20\,keV unabsorbed flux in units of $10^{-11}$ erg.s$^{-1}$\,cm$^{-2}$, power law index, iron energy in units of keV and reduced $\chi^2$ as a function of observation time. Errors indicate $1\sigma$ confidence level.}
\label{fig:specev}
\end{figure}

In order to investigate the variability of the spectrum through the
outburst of XMMU~J054134.7$-$682550, the best fit model above ({\tt power*highecut}) was
fit to all observations in Table~\ref{tab:obs} in 3--20\,keV band using a fixed $n_H$ value of $8\times 10^{20}$\,$cm^{-2}$.  For most observations
an Fe line at $\sim 6.4$\,keV was required to fit the spectra. From the spectral fits cut-off energy values were found to be between $\sim 11$ and $\sim 21$ keV, and $E_{\rm fold}$ energy values were found to be between $\sim 4$ and $\sim 33$ keV. In Fig.~\ref{fig:specev}, we plot 3--20\,keV unabsorbed X-ray flux, spectral parameters and reduced $\chi^2$ obtained from spectral fits as a function of observation days.

Assuming a distance of 49\,kpc (Catelan, Cortes, 2008) and using the unabsorbed X-ray flux values plotted in Fig.~\ref{fig:specev} and listed in Table~\ref{tab:obs}, we found that X-ray luminosity of the source in the 3--20\,keV band varies between $\sim2\times 10^{36}-9\times 10^{37}$\,erg\,s$^{-1}$.  

\begin{figure}
\includegraphics[width=75mm]{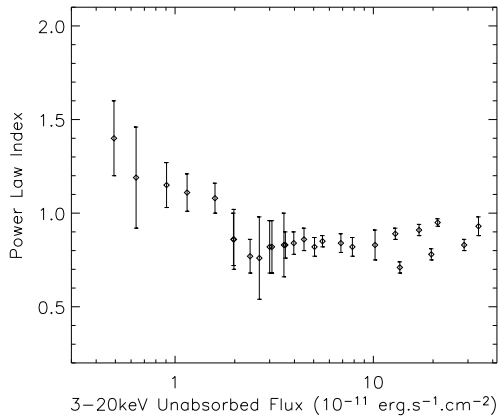}
\caption{Variation of power law index with X-ray flux.}
\label{fig:speccorrelations}
\end{figure}

Fig.~\ref{fig:speccorrelations} shows the dependence of power law index on the flux. The power law index shows an anticorrelation with the flux, i.e. the spectrum gets softer with decreasing X-ray flux. Spectral softening accompanied with X-ray flux decrease can be considered to be a result mass accretion rate variations (Meszaros et al. 1983, Harding et al. 1984), and is also observed in other accretion powered X-ray pulsars like SAX~J2103.5+4545 (Baykal et al. 2002) and 2S~1417$-$62 (Inam et al. 2004). 

\section{Pulse Phase Resolved Spectroscopy}

As cyclotron features are strongly dependent on viewing angle, phase resolved spectroscopy was undertaken to further study the presence of possible cyclotron features.  A secondary objective was to observe whether any other spectral features changed with the phase of the neutron star spin. Because of the distance to the source and the weakness of the possible cyclotron feature, we focus here on the brightest of the observations (01-00 at MJD 54321.1) to ensure the best signal-to-noise possible. 

\begin{figure}
\includegraphics[width=65mm,angle=90]{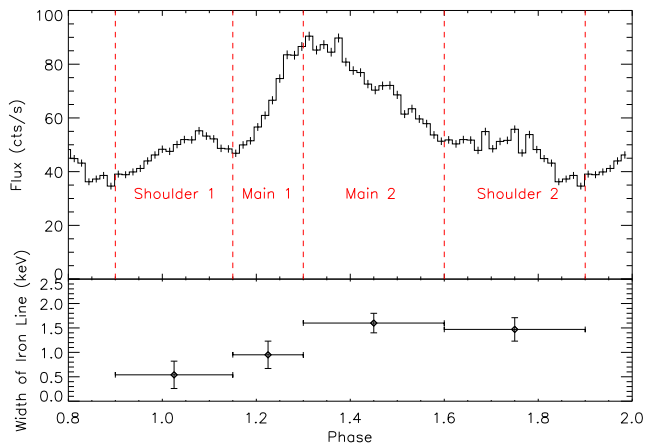}
\caption{Upper Panel: Pulse profile for observation  01-00. The four phases studied are marked. Lower Panel: Width of iron line as a function of pulse phase using the power law model with a high energy cutoff. The y-errors are the 90\% uncertainties given by XSPEC.}
\label{fig:phaseres}
\end{figure}

The neutron star pulse period was arbitrarily divided into four phases according to the general morphology of the pulse profile, and in such as way that the chosen phases had broadly adequate signal-to-noise.  The upper panel of Fig.~\ref{fig:phaseres} shows the pulse profile for observation 01-00 folded on a 61.6\,s period with the phase bins comprising two shoulder features and the rising and falling limbs of the main peak.

Using the spectral fits to the phase averaged spectrum in the previous section as a guide, we fitted a cutoff power law model to the phase resolved spectra to see if there were any significant deviations from the phase averaged continuum. The hydrogen column density was fixed to the standard LMC value of $2.5 \times 10^{21} \rm{cm}^{-2}$ (Kalberla et al. 2005).  A simple absorbed cutoff power law was an adequate fit ($\chi^2_\nu<1.1$) to the data for the Shoulder 1 and Main 1 phases, but an iron line at 6.4\,keV (fixed) was required in the Main 2 and Shoulder 2 spectra. There was found to be little variation in the slope of the spectra, and all four showed weak signs of an absorption feature at 10\,keV, much the same as the phase averaged spectrum.  In the case of the Main 2 spectrum, however, the $\chi^2_\nu$ was improved from 1.3 to 1.1 by the addition of a cyclotron feature at 10\,keV.  The F-test probability of this improvement being due to chance is 0.007, suggesting a rather marginal detection.  

To be consistent with the analysis of the phase average spectra, we also fitted an absorbed power law model with a high energy cutoff. Again, this proved similar to the phase averaged spectra in that the cutoff in the power law masked the residuals at 10\,keV that were apparent in fitting the cutoff power law above.  Once again, the hydrogen column density was fixed, and the energy of the iron line was fixed at 6.4\,keV.  With this model, too, the spectral slope showed little variation with pulse phase and no cyclotron features were required.  
Table~\ref{tab:phaseres} gives the parameters for the fits to the four spectra. 

\begin{table*}
\caption{Parameters from the powerlaw model with a high energy cutoff fitted to the four phases indicated in Fig~\ref{fig:phaseres}.  The centroid of the iron line was fixed at 6.4\,keV and the $N_H$ at $2.5\times10^{21}$\,cm$^{-2}$ in all cases.  Errors are 90\% uncertainties.}

%\begin{center}

\begin{tabular}{l c  c  c  c } \hline
Parameter  &  Shoulder 1 & Main 1 & Main 2 & Shoulder 2  \\ \hline
%$N_H$ ($10^{22}{\rm cm}^{-2}$) (fixed) & $0.25$ & $0.25$ & $0.25$ & $0.25$ \\
%Iron Line Energy (keV) (fixed) & $6.4$ & $6.4$ & $6.4$ & $6.4$ \\
$\sigma_{\rm Fe}$ (keV) & $0.5\pm 0.3$ & $1.0\pm0.3$ & $1.6\pm0.2$ & $1.5\pm0.2$ \\
$I_{\rm Fe}$ (ph\,cm$^{-2}$\,s$^{-1}$)  & $(4\pm2)\times10^{-4}$ & $(9\pm4)\times10^{-4}$ & $(1.5\pm0.5)\times10^{-3}$ & $(9.1\pm0.5)\times10^{-4}$ \\
$\Gamma$ & $0.91\pm0.02$ & $0.86\pm 0.03$ & $0.91\pm0.01$ & $0.90\pm0.02$ \\
%Iron Line Equivelent Width (eV) & $234\pm 33-402$ & $411\pm 94-745$ & $569\pm 269-881$ & $636\pm 291-1063$ \\
$\chi^2_\nu$ (58 d.o.f) & 0.69 & 0.81 & 0.92 & 1.10 \\ \hline
\end{tabular}
\label{tab:phaseres}
%\end{center}
\end{table*}

The lower panel of Figure~\ref{fig:phaseres} shows the variation in the width of the iron line for the four phases of our pulse profile for the high energy cutoff model with y-error bars the 90\% uncertainties from XSPEC. It is also clear from Table~\ref{tab:phaseres} that the iron line intensity follows the intensity of the source through the pulse profile, i.e. the iron line is strongest in the pulse peak.  In Shoulder 1 the iron line is poorly constrained, and the low $\chi^2_\nu$, together with the errors on the line width indicate that it is not strictly required in this phase.  The errors are large due to the faintness of both the iron line and are probably influenced by uncertainty in the underlying continuum. This makes it difficult to infer clearly any variation of the width of the line with phase from these data. 

Neither of the models fitted point strongly towards the presence of a cyclotron feature in this source, however, given the quality of the data, a cyclotron line cannot be ruled out.  The data suggest that there may be some variation in both the width and intensity of the 6.4\,keV iron line through neutron star spin phase, but further observations below 3\,keV, which will constrain the hydrogen column density, are required to show any such variation with certainty.

\section{Discussion and Conclusion}

The lack of orbital signature in pulse frequency and phase residuals is evident in Fig.~\ref{fig:frequency}.  We also searched for a possible periodicity in the RXTE-ASM lightcurve as a sign of the binary orbit, but we did not find a sign of orbital modulation. The lack of orbital signature may be due to the fact that the orbital period of the binary sytem is larger than the total observation time span of $\sim 74$ days.  Future X-ray monitoring observations are needed to detect the signs of the orbit of this system.    

The transient nature of the source may be a sign of a Be type companion, i.e. the enhancement in the accretion rate may be due to the presence of a temporary accretion disk that forms when the neutron star enters denser equatorial stellar wind of the Be companion. Thus, the outburst is probably a type II outburst from a
Be star system. From Fig.~\ref{fig:asm}, this outburst is seen to be the brightest outburst from this source since the launch of RXTE. 

In Fig.~\ref{fig:frequency}, a spin-up of trend of the source is evident especially for the first $\sim 40$ days of the observations. In Fig.~\ref{fig:fderiv}, it is seen that flux is correlated with spin-up rate of the source. From Fig,~\ref{fig:fderiv}, it is evident that observations corresponding to  the lowest flux values have spin rates consistent with zero, showing that source is either spinning-up or it is at an almost constant spin frequency phase without any clear evidence of a spin-down episode. 
If we assume that observed X-ray luminosity is proportional to the bolometric
luminosity (or mass accretion rate), then spin-up rate (which is proportional to torque exerted on the neutron star) and X-ray
flux correlation can be explained by accretion from accretion discs
when the net torque is positive and of the order of the material
torque (Ghosh \& Lamb 1979; Ghosh 1993). There are transient X-ray pulsar systems which show correlation between spin-up rate and X-ray flux:
EXO~2030+375 (Parmar, White \& Stella 1989; Wilson et al. 2002),
A~0535+26 (Finger,Wilson \& Harmon 1996a; Bildsten et al. 1997),
2S~1845$-$024 (Finger et al. 1999), GRO~J1744$-$28 (Bildsten et al.
1997), GRO~J1750$-$27 (Scott et al. 1997), XTE~J1543+568 (in't
Zand, Corbet \& Marshall 2001) and SAX~J2103.5+4545 (Baykal,
Stark \& Swank 2002, Baykal et al. 2007), 2S~1417$-$62 (Finger et al. 1996b, Inam et al. 2004). Thus, the positive correlation we show in Fig.~\ref{fig:fderiv} may be an indication of accretion via an accretion disk. However, without any orbital correction, it is not possible to analyze the nature of this correlation quantitatively, since the calculated pulse frequencies might not accurately correspond to the pulsar's spin frequencies.

Decreasing pulsed fraction with decreasing X-ray flux (Fig.~\ref{fig:pfrac}) and changing pulse morphology (Fig.~\ref{fig:profiles}) may indicate changes in accretion geometry and even be an indicator of a transition from disc to wind accretion, or an ongoing transition from accretor phase to propeller phase. This is similar to the decrease in pulsed fraction and pulse morphology changes in the low X-ray flux observations in 2S~1417$-$62 (Inam et al. 2004). The evidence that observations corresponding to the lowest flux values have spin rates consistent with zero (Fig.~\ref{fig:fderiv}) may also be an indicator of accretion geometry changes in the source.

The decline of the X-ray flux together with the change in pulse profile shapes and pulsed fraction may be an indication of a transition from accretor to propeller stage. When the propeller stage sets in, the great majority of accreting matter cannot reach the neutron star. To test this hypothesis, we can compare the factor of decrease in the bolometric luminosity with the theoretical expectation for accretion to propeller stage, 
 
\begin{equation}
\Delta=170M_{1.4}^{1/3}R_6^{-1}P_0^{2/3}
\end{equation}

where $M_{1.4}$ is the mass of the neutron star in units of $1.4M_{\odot}$, $R_6$ is the radius of the neutron star in units of $10^6$ cm and $P_0$ is the spin
period of the neutron star in units of seconds (Corbet 1996; Campana
\& Stella 2000; Campana et al. 2002). The factor $\Delta$ becomes $\sim 2.6\times 10^4$
for a neutron star with mass $\sim 1.4M_{\odot}$, radius $\sim10^6$ cm and spin
period of $\sim61$\,s. If we assume that the factor of decrease in bolometric luminosity is comparable to that of 3--20\,keV luminosity, we find that our observed factor of luminosity change is $\sim 50$, which is much less than the theoretical expectation for a transition to propeller stage. It is not likely
that we observe any transition to propeller stage.

As seen in Fig.~\ref{fig:profiles}, pulse profiles show high variability. Except 3--8, 8--13 and 13--18\,keV pulse profiles of the observation on MJD 54321.1 and 18--25\,keV pulse profile of the observation on MJD 54351.6, pulse profiles have two peaks seperated by $\sim 0.5$ phase. 

We found that both the phase average spectrum could be fit with either a cutoffpl model, with a possible weak cyclotron feature at 10\,keV, or that it could be fit adequately using a power law with high energy cut-off model without the need for a cyclotron feature.  We used the power law model with high energy cutoff to map the spectral evolution throughout all observations (see Fig.~\ref{fig:specev}).  

We also constructed pulse phase resolved spectra from the brightest observation. From this analysis (see Fig.~\ref{fig:phaseres}), we did not find any conclusive evidence of a cyclotron absorption feature, possibly due to low count statistics of the source. We found a very marignal variation in the iron line width and intensity which will need to be confirmed through further observations in the soft X-ray band.

{\bf{Acknowledgments}}

S.\c{C}. \.{I}nam and A. Baykal acknowledge support from T\"{U}B\.{I}TAK, the Scientific and Technological Research Council of Turkey through project 106T040 and EU FP6 Transfer of Knowledge Project "Astrophysics of Neutron Stars"
(MTKD-CT-2006-042722). We would like to thank Jorn Wilms and the IAAT library for the IDL routines used in this analysis. L. Townsend is funded by a Mayflower grant and would like to thank the School of Physics and Astronomy for their support.

\noindent{{\bf{References}}}

Baykal A., Stark M., Swank J., 2002, ApJ, 569, 903

Baykal A., Inam S.C., Stark M.J., Heffner C.M., Erkoca A.E., Swank J.H., 2007, MNRAS, 374, 1108 

Bildsten L. et al., 1997, ApJS, 113, 367

Campana L., Stella L., 2000, ApJ, 541, 849

Campana S., Stella L., Israel G.L., Moretti A., Parmar A.N., Orlandini M.,
2002, ApJ, 580, 389

Corbet R.H.D., 1996, ApJ, 457, L31

Deeter J.E., Boynton P.E., 1985, in Hayakawa S., Nagase F., eds, Proc.
Inuyama Workshop on Timing Studies of X-Ray Sources. Nagoya Univ.,
Nagoya, p.29

Finger M.H., Wilson R.B., Harmon B. A., 1996a, ApJ, 459, 288

Finger M.H., Wilson R.B., Chakrabarty D., 1996b, A\&ASS, 120, 209

Finger M.H., Bildsten L., Chakrabarty D., Prince T.A., Scott D.M.,Wilson
C.A., Wilson R.B., Zhang S.N., 1999, ApJ, 517, 449

Ghosh P., 1993, in Holt S.S., Day C.S., eds, The Evolution of X-ray Binaries.
Am. Inst. Phys., New York, p. 439

Ghosh P., Lamb F.K., 1979, ApJ, 234, 296

Harding A. K., Kirk J. G., Galloway D. J., Meszaros P., 1984, ApJ, 278, 369

Inam S.C., Baykal A., Scott, D.M., Finger M., Swank J., 2004, MNRAS, 349, 173

Jahoda K., Markwardt C.B., Radeva Y., Rots A.H., Stark M.J., Swank J.H., Strohmayer T., Zhang W., 2006, ApJS, 163, 401

in't Zand J.J.M., Corbet R.H.D., Marshall F.E., 2001, ApJ, 553, L165

Kalberla et al. 2005 A\&A, 440, 775

Liu Q.Z., van Paradijs J., van den Heuvel E.P.J., 2005, A\&A, 442, 1135

Markwardt C.B., Swank J.H., Corbet R., 2007, ATEL 1176

Meszaros P., Harding A.K., Kirk J.G., Galloway D.J., 1983, ApJ, 266,
L33

Palmer D.M., Grupe D. \& Krimm H.A., 2007, ATEL 1169

Parmar A.N., White N.E., Stella L., 1989, ApJ, 184, 271

Scott D.M., Finger M.H., Wilson R.B., Koh D.T., Prince T.A., Vaughan
B.A., Chakrabarty D., 1997, ApJ, 488, 831

Shtykovskiy P., Gilfanov M., 2005, A\&A, 431, 597

White N.E., Swank J.H., Holt S.S., 1983, ApJ, 270, 711 

Wilson C.A., Finger M.H., Coe M.J., Laycock S., Fabregat J., 2002, ApJ,
570, 287

\end{document}